\documentclass[12pt]{revtex4}
\usepackage{graphicx,amsmath,amssymb}

\def\bc{\begin{center}}
\def\ec{\end{center}}
\def\beq{\begin{equation}}
\def\eeq{\end{equation}}

\begin{document}

\title{Two-Jet Inclusive Cross Sections in Heavy-Ion
Collisions in the Perturbative QCD}

\author{M.A. Braun$^a$, R.S. Kolevatov$^a$, B. Vlahovic$^b$}
\affiliation{$^a$S.Petersburg State University, Russia, $^b$ North
Carolina Central University, NC, USA}

%\begin{center}
%{\Large\bf Two-Jet Inclusive Cross Sections in Heavy-Ion Collisions
%in the Perturbative QCD}
%\vskip 0.5 cm
%M.A. Braun$^a$, R.S. Kolevatov$^a$, B. Vlahovic$^b$
% {\it $^a$S.Petersburg State University, Russia, $^b$North Carolina
%Central University, NC, USA}
%\end{center}
%\vskip 0.5 cm

\begin{abstract}
In the framework of perturbative QCD, double inclusive cross
sections for high $p_t$ parton
%jet
production in nucleus-nucleus collisions are
computed with multiple rescattering taken into account. The
induced long-range correlations between
%%% RK
numbers of jets at forward and backward rapidities
%%%%%%%%
% jet multiplicities
are found to reach $10\div 20$\% for light nuclei at
$\sqrt{s}=200$~GeV/c and to be suppressed for heavy nuclei and LHC
energies.
\end{abstract}

\maketitle

\section{Introduction}
Particle production in high-energy heavy ion collisions is now at
the center of experimental efforts to discover the quark-gluon
plasma \cite{qgpRHIC,qgpNA49}. The observed particle spectra are
the result of different mechanisms which are responsible for the
creation of initial
%%%%% RK %%%%%%
high-$p_t$ partons,
%%%%%%%%%%%
%jets,
%
their propagation and subsequent hadronization. (Experimentally
the initially produced partons can be traced as jets of hadrons,
so that in the following we shall often use the term 'jet' for the
produced parton in the pure theoretical sense, neglecting all the
subtleties related to its actual determination in the experiment.)
To be able to see the formation of the quark-gluon plasma against
the background of more conventional effects, such as
 as gluon emission due to bremsstrahlung and
multiple hard collisions in the nuclear medium, one has to fully
study the consequences of the latter. Jet quenching due to gluon
emission has been studied in considerable detail (see e.g.
~\cite{bai,sal}). Also the effects
of %multiple  hard scattering %
rescattering in single particle inclusive spectra
%RK
initiated by the observation of \cite{Cronin}
has been considered in a series of articles for multiple soft
\cite{mults,Kop} and multiple hard pQCD scatterings
~\cite{oldhard,calu,trel1,bra,AcGyu,trel2}.
%%%%%%%%%%%%%%%%%%%%%%%%%%%%%%%%%%%%%%%%%%%%%%%%%%%%%%%%%%%5
This paper generalizes the study of multiple hard collisions to
double inclusive cross-sections and the following long-range
correlations between the secondaries. Observation of such
correlations has always  served  a very precise instrument for
analyzing the dynamics of the interaction.

Following the framework introduced in ~\cite{bra} we consider only
relatively hard collisions which allow for the perturbative QCD
approach. The colliding nuclei are assumed to contain a variable
number of partons   with initially small transverse momenta, which
become large as a result of hard collisions between partons
belonging to the projectile and target. The latter are assumed to
move fast along the collision axis in the opposite directions.
%%%%%%%%%%%%%%%%%
%%%RK%%%%%%%%%%%%
In the present work we restrict ourselves to the study of high
$p_t$ parton production at rapidities well separated from central
rapidity region. Here longitudinal momenta of partons in the final
state are much larger than the transverse ones. So we neglect
attenuation of the former during the collisions and assume them to
be conserved throughout the nucleus-nucleus interaction in
agreement with the standard Glauber treatment.
%%%%%%%%%%%%%%%%%
%%%%%%%%%%%%%%%%%
%Their longitudinal momenta are assumed to be much larger than the
%transverse ones and conserved during collisions, so that we neglect
%quenching at this stage of investigation.
%The multiparton
%distributions, which enter the multiple collisions formulas are assumed
%to factorize. This means that
We  neglect intrinsic correlations between partons inside the
colliding nuclei which correspond to nuclear shadowing and take
the partonic distributions just as a product of such distributions
for the nucleons smeared out with the standard nuclear profile
functions. This implies that we correctly describe quantum
evolution of these distributions except for effects coming from
interaction of partons between different nucleons in the nucleus,
which should be studied from the DGLAP evolution equation for the
nucleus as a whole or, at very small $x$, in the framework of the
BFKL evolution (e.g. in the Colour Glass Condensate approach, see
~\cite{CGC} and references therein). Jet quenching and hard parton
contents of the participants are  some of  other important
phenomena to be taken into account in the full treatment of jet
production. Thus the physical phenomenon we study is restricted to
multiple hard elastic scattering between partons. Our study is
then to be considered as a baseline calculation to be followed by
inclusion of all the above-mentioned effects

Our  formalism closely follows ~\cite{bra} to which paper we refer
the reader for details. As in ~\cite{bra}, our basic
approximations in the study of multiple partonic collisions are 1)
purely elastic partonic collisions with conservation of
longitudinal momenta and 2) factorization of the $S$-matrix into
the product of elementary partonic $S$- matrices (the Glauber
approximation). We shall also use the notations of ~\cite{bra}. To
make our presentation more self-contained we reproduce some basic
notations below.

The initial states of the colliding nuclei A and B are represented
as a superposition of states with different number of partons,
having specific values of scaling variables and transverse
positions with respect to the target nucleus center. We combine
them into a single argument $z=\{v,b\}$ for nucleus A and
$u=\{w,c\}$ for nucleus B. The (variable) number of partons will
be denoted correspondingly by $n$ and $l$. In these notations the
initial state of the nucleus  A is \beq
|A\rangle=\sum_{n=1}^{\infty} \int d\tau_A(n)
\Psi_{A,n}(z_1,\ldots,z_n)|n,z_i\rangle, \eeq where
$d\tau_A(n)=\prod\limits_{i=1}^{n}d^3z_i$ stands for the phase
space volume of the configuration. Wave functions $\Psi_{A,n}$ of
the $n$-parton configuration, symmetric in their arguments, are
normalized to fulfill $\langle A|A\rangle=1$: \beq W_n \equiv \int
d\tau(n) \left| \Psi_{A,n} (z_1,\ldots,z_n) \right|^2, \quad
\sum_n W_n = 1. \eeq We will assume a Poissonian distribution for
the number of partons: $w_n=e^{-\langle n \rangle}/n!$ and, as we
neglect intrinsic partonic correlations, a factorization property
for the wave function \beq \label{facprop} \left| \Psi_{A,n}(z_i)
\right|^2=c_n\prod_{i=1}^n \rho_A (z_i). \eeq \label{rhonorm} The
Poissonian distribution immediately implies \beq
c_n=\frac{1}{n!}e^{-\langle n \rangle}, \quad \langle n
\rangle=\int d^3z \rho_A (z), \eeq where $\langle n \rangle$
corresponds to the average number of partons in nucleus A. The
same notations are used also for nucleus B.

%%%% RK %%%%%%%%%%%%%%%%%
The paper is organized as follows. In sections \ref{samedir} and
\ref{difdir} we derive expressions for 2-parton inclusive cross
section for two different cases. The case when both partons belong
to the same forward rapidity region is examined in the section
\ref{samedir}, derivation for the forward-backward case is
presented in section \ref{difdir}. The form of the two-parton
cross sections implies that non-trivial forward-backward
correlations emerge in high-$p_t$ parton production. In section
\ref{corr} a quantity measuring the strength of correlations,
correlation coefficient, is proposed which appears to be expressed
in a simple way in terms of the derived cross sections. Numerical
values for the correlation coefficient followed by  discussion are
presented in section \ref{numres} for light and heavy nuclei
interactions at RHIC and LHC energies. The interpretation of our
results and conclusions follow in section \ref{concl}.
%%%%%%%%%%%%%%%%%%%%%%%%

\section{Two jets in the same direction}
\label{samedir}

The double inclusive cross-section to produce two jets in the same
(forward) direction is obtained as a direct generalization of the
single inclusive cross-section in ~\cite{bra}.
%%% RK %%%%%%%%%%%%%%%%%
As indicated above, here we limit our studies to the production
of partons at rapidities well separated from zero and neglect
changes in the parton longitudinal momenta during the interaction.
This implies that both observed partons originate from the same
nucleus (projectile A).
%%%%%%%%%%%%
% Conservation of longitudinal momenta
%implies that they originate from the same nucleus (projectile A ).
%%%%%%%%%%%

%%%%%%%%%%%
%As in ~\cite{bra} we combine the scaling variable $x$ and impact parameter
%$b$ for a parton from nucleus A in a single argument $z=\{x,b\}$. Likewise
%a parton from nucleus B will be characterized by its scaling variable
%and impact parameter combined in a single argument $u$. The (variable) number
%of partons in nucleus A will be denoted by $n$ and that in nucleus B by $l$.
%The wave functions of the two observed partons from nucleus A (moving along the
%$z$-axis)  will be
%denoted by $\psi_{\alpha_1}(z_1)$ and $\psi_{\alpha_2}(z_2)$, where
%$\alpha_i=\{x_i,p_i\}$ $i=1,2$ and $x$ and $p$ are the scaling variable
%and transverse momentum respectively.
%%%%%%%%%%%%%%%%%%%%%%

%%%RK%%%
We have to fix the final state $A'$ of the nucleus A to describe
the two produced partons. Since the total wave function is
symmetric in all the partons we  choose them to be the first and
the second one and account for other possibilities by factor
$\sqrt{n(n-1)}$:
%%%%%%%
%The $n$-parton wave function of the final nucleus A is given by
%%%%%%%%%%%%
\beq \label{samedirwf}
\Psi_{A',n}(z_1,z_2,z_3,...z_n)=\sqrt{n(n-1)}\psi_{\alpha_1}(z_1)
\psi_{\alpha_2}(z_2)\tilde{\Psi}_{A',n-2}(z_3,...z_n) \eeq
%%%%%%%%%%%%
% where factor $\sqrt{n(n-1)}$ takes into account the possibility to
% observe any pair of the produced partons.
%%%%%%%%%%%%%%%
%RK
where $\tilde{\Psi}_{A',n-2}(z_3,...z_n)$ is the symmetrized wave
function of the unobserved $n-2$ partons and
$\psi_{\alpha_i}(z_i)$, $i=1,2$ are the wave functions of the
observed partons in the final state. $\alpha_i$ combine scaling
variables and transverse momenta $p_1$ and $p_2$ of the latter.

The  probability to observe the two partons is given by the
modulus squared of the $S$-matrix element $\langle AB | S | A'B'
\rangle$ summed over all possible final states of the unobserved
partons.
%%%%%RK
In our approximation (purely elastic collisions) the number of
partons is not changed by the interaction. So the $S$-matrix is
diagonal in the basis $\{n,z_i; l,u_i\}$:
\begin{multline}
\langle n',z'_i,l',u'_j |S| n,z_i;l,u_j\rangle\\
=\delta_{nn'}\delta_{ll'}\prod_{i=1}^{n}\delta^{(3)}(z_i-z'_i)\prod_{j=1}^{l}\delta^{(3)}(u_i-u'_i)
S_{nl}(z1,\ldots,z_n|u_1,\ldots,u_l).
\end{multline}
%%%%%
Following ref.~\cite{bra} we take the square modulus of (6) making
use of the specific form for the final-state wave function of the
projectile nucleus (\ref{samedirwf}). The double differential
cross-section at a fixed overall impact parameter $\beta$ which
follows reads

%%%%%%
%Taking the relevant matrix element, following  ref.~\cite{bra}
%we obtain the double differential cross-section at a fixed overall
%impact parameter $\beta$ as
%%%%%
\[
\frac{d\sigma_{\alpha_1\alpha_2}}{d^2\beta}= \sum_{n=2}^\infty
\sum_{l\ge0} n(n-1)\int dz_1dz_2dz'_1dz'_2
\psi_{\alpha_1}(z'_1)\psi_{\alpha_2}(z'_1)\psi_{\alpha_1}^*(z_1)
\psi_{\alpha_2}^*(z_2)\]\[ \int
d\tau_A(n-2)d\tau_B(l)\Psi_{A,n}^*(z'_1,z'_2,z_3,...z_n)
\Psi_{A,n}(z_1,z_2,z_3,...z_n)|\Psi_{B,l}(u_1,\ldots,u_l)|^2\]\beq
\label{sigsamedir} \bigg[S_{nl}(z_1,z_2,z_3...z_n|u_1,...
u_l)-1\bigg] \bigg[S_{nl}^*(z'_1,z'_2,z_3...z_n|u_1,...
u_l)-1\bigg] \eeq

%RK%%%
The product of square brackets in (\ref{sigsamedir}) gives four
terms. The double inclusive cross-section for  transverse momenta
$p_1, p_2 \gg 1/R_{A,B}$ corresponds only to the term which
originates from the product of the $S$-matrices.
%%%%%%%%%%%%%%%%%%%%%%%%%%%%%%%%%%%%%%%%%%%%%%%%%%%%%%%%%%%%%%
Indeed in the Glauber approximation we assume that the $S$-matrix
is a product of $S$-matrices for pair parton collisions. Then
terms linear in $S$ or $S^*$ diagrammatically correspond to
cutting the forward scattering amplitude either to the extreme
left or to the extreme right of all partonic interactions. In both
cases the momenta of intermediate partons in the cut coincide with
their values in the initial colliding nuclei, so that their
transverse momenta are small, of the order of typical nuclear
scale $1/R_{A,B}$ (see ~\cite{bra} for more detail).
%As for the single inclusive cross-section, the double inclusive
%cross-section for  transverse momenta $p_1, p_2>>1/R_{A,B}$ corresponds
%only to the term which originates from the product of the $S$-matrices
%in the square brackets.
%%%RK%%%

We present each of the $S$-matrices as a product of the elementary
partonic ones, \beq
S_{nl}(z_1,\ldots,z_n|u_1,\ldots,u_l)=\prod_{i=1}^n \prod_{j=1}^l
s_{ij} \eeq Here $s_{ij}=1+ia(z_i,u_j)$ and $a(z_i,u_j)$ are the
parton-parton scattering matrix and amplitude respectively. Since
only elastic parton scattering  are considered, unitarity of the
partonic $s$-matrices means $s_{ij}s^*_{ij}=1$. So we obtain that

%Presenting each of the $S$-matrices as a product of elementary partonic ones
%and using the unitarity of the latter we obtain that
\[
S_{nl}^*(z'_1,z'_2,z_3...z_n|u_1,... u_l)
S_{nl}(z_1,z_2,z_3...z_n|u_1,... u_l)\]\beq=
\prod_{i=1,2}\prod_{j=1}^l [1+ia(z'_i,u_j)]^*[1+ia(z_i,u_j)]. \eeq
and does not depend on $z_3,...z_n$. This allows to integrate over
these variables and sum over $n$ to produce the density matrix
$\rho_A$
%
%$\rho$-matrix
%
of the nucleus A for a pair of partons:

\beq \sum_{n}n(n-1)\int
d\tau_A(n-2)\Psi_{A,n}^*(z'_1,z'_2,z_3,...z_n)
\Psi_{A,n}(z_1,z_2,z_3,...z_n)=\rho_A(z_1,z_2|z'_1,z'_2) \eeq
%%%%% RK
Factor $n(n-1)$ again accounts the possibility that any two of the
partons entering the symmetric wave function can be chosen.
Thus we get
\[
\frac{d\sigma_{\alpha_1\alpha_2}}{d^2\beta}= \sum_{l}\int
dz_1dz_2dz'_1dz'_2
\psi_{\alpha_1}(z'_1)\psi_{\alpha_2}(z'_1)\psi_{\alpha_1}^*(z_1)
\psi_{\alpha_2}^*(z_2)\rho_A(z_1,z_2|z'_1,z'_2)\]\beq \int
d\tau_B(l) |\Psi_{B,l}(u_j)|^2
\bigg\{\prod_{i=1,2}\prod_{j=1}^l[1+ia(z'_i,u_j)]^*[1+ia(z_i,u_j)]-1
\bigg\}. \eeq
%%%% RK
Making use of the factorization of the wave function
(\ref{facprop}) we have \beq
\rho_A(z_1,z_2|z'_1,z'_2)=\rho_A(z_1|z'_1)\rho_A(z_2|z'_2)\quad
\mbox{and}\quad
\left|\Psi_{B,l}(u_j)\right|^2=\frac{1}{l!}e^{-\langle l
\rangle}\prod_{j=1}^l\rho_B(u_j|u_j). \eeq Assuming factorization
of partonic transverse and longitudinal degrees of freedom, for
equal scaling variables $v_i=v'_i=v$, we have \beq \label{rhofact}
\rho_A(v,b_i|b'_i)=P_A(v)\tilde \rho_A(b_i|b'_i), \eeq where
$P_A(v)$ is the mean parton number distribution and $\tilde
\rho_A(b_i|b'_i)$ is the transverse part of single parton density
matrix. For equal arguments it goes into the standard profile
function of
 the projectile nucleus
$\tilde \rho_A(b_i|b_i) = T_A(b_i-\beta)$ (recall that the origin
in the transverse plane is in the center of the target nucleus B).
Similarly for the target nucleus $\rho_B(u_j|u_j)=P_B(w_j)
T_B(c_j)$.

As a result of this factorization the inclusive cross section
transforms into
%%%%%%
%Using more explicit notation, double inclusive cross section transforms to
%, using factorization of the wave function
%and $\rho$-matrix,

%%%%% RK %%%  I changed \rho_A(b_i-\beta|b'_i-\beta) into
%\tilde \rho_A(b_i|b'_i), the comments are above;
%%impact parameter enters a bit further when the density matrix transforms into the profile functions.
%%%%%%%%%%%%%%%%%%%
\[
I_{AA}(\beta,y_1p_1,y_2,p_2)\equiv \left.
\frac{(2\pi)^4d\sigma}{dy_1dy_2d^2p_1d^2p_2d^2\beta}\right|_{y_1,y_2>0}\]\[=
\sum_l\frac{1}{l!}e^{-\langle l\rangle}
\int\prod_{i=1,2}\bigg(d^2b_id^2b'_ie^{ip_i(b_i-b'_i)} P_A(v_i)
%%%%%%%
%\rho_A(b_i-\beta|b'_i-\beta)
%%%%
\tilde\rho_A(b_i|b'_i)
%%%
\bigg)\]\beq \int \prod_{j=1}^ld^2c_jdw_jP_B(w_j)
%%% RK missing factor of T_B(c_j) added
T_B(c_j)
\bigg\{\prod_{i=1,2}\prod_{j=1}^l[1+ia(z'_i,u_j)]^*[1+ia(z_i,u_j)]-1
\bigg\} \eeq where the rapidities $y_1$ and $y_2$ correspond to
the scaling variables $v_1$ and $v_2$. The subscript $AA$ for the
inclusive cross-section indicates that both jets are produced from
the nucleus $A$, although the collision is between A and B.

The non-trivial integral over $u_j=\{w_j,c_j\}$ factorizes to give
the $l$-th power of \beq J(v_1,v_2,b_1,b'_1,b_2,b'_2)= \int
d^2cdwT_B(c)P_B(w)
\prod_{i=1,2}[1+ia(v_i,w;b'_i-c)]^*[1+ia(v_i,w,b_i-c)] \eeq so
that after summation over $l$ we get \beq
I_{AA}(\beta,y_1p_1,y_2,p_2)=
\int\prod_{i=1,2}\bigg(d^2b_id^2b'_ie^{ip_i(b_i-b'_i)}
%
%\rho_A(b_i-\beta|b'_i-\beta)
%
\tilde\rho_A(b_i|b'_i)
\bigg)\bigg(e^{J(v_1,v_2,b_1,b'_1,b_2,b'_2)-\langle
l\rangle}-1\bigg) \eeq

We are left with a problem of calculation $J$, which is now more
complicated than for the single inclusive cross-section. From  the
16 terms in the product in (7) we separate a part
\[ 1+\sum_{i=1,2}\bigg([ia(v_i,w;b'_i-c)]^*+
ia(v_i,w;b_i-c)+[ia(v_i,w;b'_i-c)]^*ia(v_i,w;b_i-c)\bigg) \]
Physically it corresponds to the case when either their is no
partonic interaction at all or only one parton from nucleus A
interacts with a given parton from nucleus B. Integration over $c$
and $w$ of this part repeats that for the single inclusive case
and gives \beq J^{(1)}=\langle
l\rangle+\sum_{i=1,2}T_B\left(\frac{1}{2}(b_i+b'_i) \right) \bigg(
F_B(v_i,b'_i-b_i)-F_B(v_i,0)\bigg) \eeq where \beq F_B(x,b)=\int
dw P_B(w)\int d^2p\frac{d\sigma(x,w)}{d^2p}e^{ipb} \eeq Putting
(9) into (8) we find that the corresponding part of the double
inclusive cross-section factorizes into a product of two single
inclusive ones: \beq \label{sigAA}
I_{AA}^{(1)}(\beta,y_1,p_1,y_2,p_2)= I_{A}(y_1,p_1) I_{A}(y_2,p_2)
\eeq where $I_A(y,p)$ is the inclusive cross-section to produce a
parton from nucleus $A$ in its collision with nucleus B at impact
parameter $\beta$ (the latter dependence implicit): \beq
I_A(y,p)=\frac{(2\pi)^2d\sigma}{dyd^2pd^2\beta}= P_A(v)\int
d^2bd^2rT_A(b-\beta) e^{ipr}
\bigg(e^{T_B(b)(F_b(v,r)-F_B(v,0))}-e^{-T_B(b)F_B(v,0)}\bigg) \eeq
So the part $I_{AA}^{(1)}$ corresponds to independent production
of the two partons as expected.

The rest 9 terms correspond to the case when both the observed
partons interact with the same parton from nucleus B. Evidently in
this case the two observed partons have to be located close to
each other, since the partonic interactions are short-ranged. As a
result we shall find under a single integral over the impact
parameter $b$ a square of the profile function $T^2(b)$. Thus this
part of the inclusive cross-section will have the order $\sim 1/
A^{2/3}$ as compared to the independent production part. Having in
mind calculation of correlations, we can neglect this part in the
first approximation, as we shall see in the following.

%%%%%%%%%%%%%%%%%%%%%%%%%%%%%%%%%%%%%%%%%%%%%%%%%%%%%%%%%%%%%%%%%%%%%%%%
%%%%%%%%%%%%%%%%%%%%%%%%%%%%%%%%%%%%%%%%%%%%%%%%%%%%%%%%%%%%%%%%%%%%%%%%

\section{Two jets in opposite directions}

\label{difdir}

In this case the wave functions of the observed partons
$\psi_{\alpha_1,\alpha_2}$ from nucleus $A$ (moving along the
$z$-axis) and nucleus $B$ (moving in opposite direction) will be
denoted by $\psi_p(z)$ and $\psi_q(u)$ respectively, where $p$ and
$q$ combine their scaling variables ($v$ and $w$ respectively) and
transverse momenta. The $n$-particle wave function for the final
state of the nucleus $A$  takes  the  form:
$$\Psi_{A',n}(z_1,...,z_n)=
\sqrt{n}\psi_p(z_1)\tilde{\Psi}_{A',n-1}(z_2,..,z_n)$$ and
similarly for the nucleus B.

Again, writing down the relevant matrix element for the transition
amplitude and again following  ~\cite{bra}, we obtain the double
inclusive cross section at a fixed overall impact parameter
$\beta$
\[I_{AB}(\beta,y_1,p,y_2,q)= \sum_{nln'l'\ge 1}
\sqrt{nn'll'}\sum_{A'B'}\int d\tau'_A(n')
d\tau'_B(l')d\tau_A(n)d\tau_B(l)
\]\[
\psi_p(z'_1)\psi_q(u'_1) \psi^*_p(z_1) \psi^*_q(u_1)
\Psi^*_{A,n'}(z'_1...z'_n)\Psi^*_{B,l'}(u'_1...u'_l)
 \tilde{\Psi}_{A,n-1}(z'_2...z'_n)
\tilde{\Psi}_{B,l-1}(u'_2...u'_l)\]\[
\tilde{\Psi}^*_{A,n-1}(z_2...z_n)
\tilde{\Psi}^*_{B,l-1}(u_2...u_l)
\Psi_{A,n}(z_1...z_n)\Psi_{B,l}(u_1...u_l)\]\beq
[S_{nl}^*(z'_1...z'_n|u'_1...u'_l)-1][S_{nl}(z_1...z_n|u_1...u_l)-1]
\eeq \noindent where the set $z_i=\{v_i,b_i\}$ of longitudinal
momenta and transverse positions corresponds to the nucleus A,
whereas the set of $u_j=\{w_j,c_j\}$ to nucleus B. The subscript
$AB$ indicates that the two jets are now produced from different
nuclei.

Summing over all unobserved states of final nuclei $A'$ and $B'$,
we can perform integrations over $z_2'...z_n'$ and $u'_2..u'_n$ to
obtain

\[I_{AB}(\beta,y_1,p,y_2,q)= \sum_{nl\ge 1}
n l\int dz_1 dz_1' d\tau_A(n-1) du_1 du_1' d\tau_B(l-1)\]
\[
\Psi^*_{A, n}(z'_1,z_2,...z_n)\Psi^*_{B, l}(u'_1,u_2...u_l)
[S^*(z'_1,z_2...z_n,u'_1,u_2,...u_l)-1] \psi_p(z'_1) \psi_q(u'_1)
\]\beq
\psi^*_p(z_1) \psi^*_q(u_1) [S(z_1...z_n,u_1...u_l)-1] \Psi_{A,
n}(z_1...z_n) \Psi_{B, l}(u_1...u_l) \eeq

The product of the amplitude and its conjugate gives three terms
\beq [S^*-1][S-1]=[S^* S-1]-[S^*-1]-[S-1] \label{ampl}\eeq For the
same reason as for the case of two jets in the same direction
discussed in the previous section,
% As
%mentioned, we assume that the $S$-matrix is a product of
%$S$-matrices for pair parton collisions. Then the second and
%third terms diagrammatically correspond to cutting the forward
%scattering amplitude either to the extreme left or to the extreme
%right of all partonic interactions. In both cases the momenta of
%intermediate partons in the cut coincide with their values in the
%initial colliding nuclei, so that their transverse momenta are
%small. Therefore
%
the second and third terms in (\ref{ampl}) give the partonic
spectrum only at very small values of transverse momenta and can
be neglected.   To be more explicit, consider terms with no more
than one partonic collision with the projectile and/or target.
Inserting transverse parts of the wave functions of observed
particles in the form $\psi \sim e^{i  k b}$ we arrive at the
integral(for the term with $S$)
\[
\int d^2b_1 d^2c d^2c_1 d^2b d^2c'_1 d^2b_1' e^{i  p b_1+i  q
c_1-i  p b'_1-i  q
  c'_1}X(b,b_1,c,c_1)
\]\beq  \Psi_A(b_1..b..)\Psi_B(c_1..c..)
  \Psi^*_A(b'_1..b..) \Psi^*_B(c'_1..c..)
  \eeq
where \beq X=ia(b_1-c)ia(c_1-b),\ \ {\rm or}\ \ ia(b_1-c),\ \ {\rm
or}\ \ ia(b-c_1) \eeq The scale at which the wave functions of
nuclei change is much greater than $1/p$ or $1/q$ so all of these
terms after integrations in $b_1',c_1'$ give contributions
proportional to the product of delta-functions
$\delta(p)\delta(q)$. If we are interested in production of
particles with large transverse momenta all these contributions
can be neglected, so that in (\ref{ampl}) only the first term
contributes.

We present the total $S$-matrix
% RK
elements
as
$$S_{nl}(z_1..z_n|u_1..u_l)=\prod_{j=1}^l s_{1j}\prod_{i=2}^n s_{i1}
\prod_ {k,m=2}^{n,l} s_{km}.$$
% RK -- it was already introduced in the same-side considerations
%where $s_{ij}=1+i  a(z_i,u_j)$ are
%parton scattering amplitudes, and $s_{ij}s^*_{ij}=1$ if we
%consider only elastic parton scattering.
%
Due to unitarity of the partonic $s$-matrix, $s_{ij}s^*_{ij}=1$,
the product $S^*S$ takes the form
\[S_{nl}^*(z'_1,z_2..z_n|u'_1,u_2..u_l)S_{nl}(z_1,z_2..z_n|u_1,u_2..u_l)=(1+i
a(z_1',u_1'))^* (1+i  a(z_1,u_1))\] \beq \prod_{j=2}^l (1+ i
a(z_1',u_j))^* \prod_{i=2}^n (1+ i  a(z_i,u_1'))^* \prod_{j=2}^l
(1+ i  a(z_1,u_j)) \prod_{i=2}^n (1+ i  a(z_i,u_1))
\end{equation}

To move further, in contrast to the derivation in ~\cite{bra}, we
are forced  from the start to make the assumption of factorization
of the nuclear wave functions. \beq
\Psi^*_A(z_1',z_2...z_n)\Psi_A(z_1,z_2...z_n)=
\rho_A(z_1'|z_1)\rho_A(z_2|z_2)..\rho_A(z_n|z_n) \frac{e^{-\langle
n \rangle}}{n!} \eeq and similarly for nucleus B,
% RK %%%%%%%%%%%
where the density matrices $\rho_A(z_i|z_i')$ have the
factorization property (\ref{rhofact}) and normalization
(\ref{rhonorm}).
%%%%
% RK -- this has already been stated in the introduction and same-side considerations.
%%%%%%%
%
%Note that for equal arguments
%\beq
%\rho_A(z|z)=P(x)T_A(b); \ \int dz \rho_A(z)=\langle n
%\rangle
%\eeq
% where $\langle n \rangle$ is the average number of
% partons in nucleus A and $T_A$ -the nuclear density and $P(x)$ is
% the parton density in longitudinal momentum space.
%

Assuming for simplicity a central collision (in other case the
argument in all
%
%RK T_B replaced by T_A to have correspondence with same-side considerations.
%
$T_A$'s must be shifted by the impact parameter $\beta$) we get
\[ I_{AB}(\beta,y_1,p,y_2,q)= \sum_{nl\ge 1} n l\int
dz_1 dz_1' du_1 du_1' \psi_p(z'_1) \psi_q(u'_1) \psi^*_p(z_1)
\psi^*_q(u_1) \rho_A(z_1|z_1')\rho_B(u_1|u_1') \]
$$
\bigg\{ [1+i  a(z_1',u_1')]^*[1+i  a(z_1,u_1)] \bigg(
\prod_{j=2}^l\int du_j \rho_B(u_j|u_j)[1+i a(z_1',u_j)]^*[1+i
a(z_1,u_j)] \bigg)
$$
\beq \bigg( \prod_{i=2}^n\int dz_i \rho_A(z_i|z_i)[1+i
a(z_i,u_1')]^*[1+i  a(z_i,u_1)]\bigg) - \int \prod \rho_B \prod
\rho_A \bigg\} \frac{e^{-\langle n \rangle-\langle l \rangle}}{l!
n!} \eeq

Already at this stage it is convenient to separate from the total
inclusive cross-section its part which does not contain
interactions between the observed partons, that is the term which
comes with unity from the product of the first two square brackets
in (28). Comparing with \cite{bra} we find it to be:
\[ I_{AB}^{(1)}(\beta,y_1,p,y_2,q)= \bigg\{\sum_{n\ge 1}
\frac{e^{-\langle n\rangle}}{(n-1)!} \int dz_1 dz_1'  \psi_p(z'_1)
\psi^*_p(z_1) \rho_A(z_1|z_1')
\]\[
 \bigg(
\prod_{j=2}^l\int du_j \rho_B(u_j|u_j)[1+i a(z_1',u_j)]^*[1+i
a(z_1,u_j)] - \int\prod_{j=2}^ldu_j\rho_B(u_j|u_j)\bigg)\bigg\}
\]\[
\bigg\{\sum_{l\ge 1}\frac{e^{-\langle l\rangle}}{(l-1)!} \int
du_1du'_1\psi_q^*(u_1)\psi_q(u'_1)\rho_B(u_1|u'_1)\]\[ \bigg(
\prod_{i=2}^n\int dz_i \rho_A(z_i|z_i)[1+i a(z_i,u_1')]^*[1+i
a(z_i,u_1)] - \int\prod_{i=2}^n dz_i\rho_A(z_i|z_i)
\bigg)\bigg\}\]\beq =I_A(y_1,p)I_B(y_2,q) \eeq where $I_{A(B)}$ is
the inclusive cross-section to produce a single jet from nucleus
$A(B)$ in AB collisions at impact parameter $\beta$ (see (20)). So
this term factorizes into a product of two independent single
inclusive cross-sections to produce each of the observed partons.
It is important that this result is exact and not based on the
smallness of the parton interaction range on the nuclear scale. In
particular, its validity is not spoiled by corrections of the
order $1/A^{2/3}$.

Now we turn to the rest terms in (28). In them the integrals in
round brackets can be done exactly as in the standard Glauber
derivation, taking into account that the space range of the
partonic interaction is much smaller than the nuclear scale (see
~\cite{bra}). After that summations in $n$ and $l$ can be easily
performed and we get
\[ I^{(2)}_{AB}(\beta,y_1,p,y_2,q)=
\int d^2b_1 d^2 b_1' d^2 c_1 d^2 c_1'
\]\[
e^{i p(b_1-b_1')+i  q(c_1-c_1')} \rho_A(v,b_1|b_1')
\rho_B(w,c_1|c_1')\]\beq \bigg\{\bigg([1+i  a (b_1'-c_1',v,w)]^*
[1+i a(b_1-c_1,v,w)]-1\bigg)e^{E_B(v,b_1,b'_1)+E_A(w,c_1,c'_1)} -1
\bigg\} \eeq where \beq
E_B(v,b_1,b'_1,)=T_B((b_1+b_1')/2)(F_B(v,b_1-b_1')-F_B(v,0)) \eeq
$E_A$ is defined by a similar formula for nucleus A and $y_1,\
y_2\ ,p_1,\ p_2$ are the rapidities and transverse momenta of the
observed particles(partons), and $v,w$ are longitudinal momentum
fractions corresponding to $y_1$ and $y_2$ respectively. Functions
$F_{B}(v,b)$ is the Fourier transform of the transverse momentum
distributions $I(v,w,p)$ in the elastic scattering of two partons
with scaling variable $v$ and $w$, averaged over the longitudinal
momenta of nucleus B partons (see ~\cite{bra}): \beq F_B(v,b)=\int
dw P(w)\int\frac{d^2p}{(2\pi)^2}I(p,v,w)e^{ipb} \eeq

It is convenient to introduce new variables for integration.
Define $$r_1\equiv b_1-b_1'; \ r_2\equiv c_1-c_1'; \ b\equiv
(b_1+b_1')/2;\ \ c\equiv (c_1+c_1')/2$$
%and then
%$$x\equiv (b+c)/2;\ y\equiv b-c$$
In these variables the cross section is rewritten as
\[ I^{(2)}_{AB}(\beta,y_1,p,y_2,q)=
\int d^2r_1d^2r_2d^2bd^2c e^{i  p r_1+i  q r_2} \rho_A(v,b_1|b_1')
\rho_B(w,c_1|c_1')
\]\beq
 \bigg\{\bigg([1+i  a
(b-c-\frac{r_1-r_2}{2},v,w)]^* [1+i  a(b-c+\frac{r_1-r_2}{2},v,w)]
-1\bigg) e^{E_B(v,b,r_1)+E_A(w,c,r_2)}-1 \bigg\} \eeq where in the
new variables \beq E_B(v,b,r_1)= T_B(b)(F_B(v,r_1)-F_B(v,0)) \eeq
and similarly for $E_A$.

Assuming that $E(..b,c...)$ and density matrices change
significantly only when $b$ and $c$  suffer macroscopic shifts of
about nucleus radius, we can perform integrations in $c$. Take the
terms in (20) containing single amplitudes $ia(b-c\pm
(r_1-r_2)/2)$. On the nuclear scale they can be effectively
substituted as
\[ia(b-c\pm (r_1-r_2)/2)\to i\tilde{a}(0)\delta^2(b-c\pm (r_1-r_2)/2)
\simeq i\tilde{a}(0)\delta^2(b-c)\] where $\tilde{a}$ is the
amplitude in the transverse momentum space and we have used that
$(r_1-r_2)/2$ is small on the nuclear scale. In the term with the
product
\[a(b-c+(r_1-r_2)/2)a^*(b-c- (r_1-r_2)/2) \]
we pass from the integration variable $c$ to $r=b-c+(r_1-r_2)/2$
in which this product takes the form
\[a(r)a^*(r- r_1+r_2) \]
Obviously $r$ is also small on the nuclear scale, so that $c\simeq
b$ and we may take all the $c$-dependent function out of the
integral over $r$ at $c=b$. The integral over $r$ takes the form
\[ \int d^2ra(r)a^*(r- r_1+r_2)=
\int d^2pe^{ip(r_1-r_2)}\frac{d\sigma(p,v,w)}{d^2p}\]

In this way we finally get for the second part of the inclusive
cross-section

\[
 I^{(2)}_{AB}(\beta,y_1,p,y_2,q)=P_A(v)P_B(w)\int d^2bT_A(b)T_B(b)
\int d^2r_1d^2r_2e^{ipr_1+iqr_2}\]\[ \bigg(\int d^2l
e^{il(r_1-r_2)}\frac{d\sigma(l,v,w)}{d^2l}
-\sigma^{tot}(v,w)\bigg)\] \beq
\bigg(e^{T_A(b)(F_A(w,r_2)-F_A(w,0))+T_B(b)(F_B(v,r_1)-F_B(v,0))}
-1\bigg) \eeq

This part of the cross-section corresponds to the case when the
two observed partons interact with each other. Obviously  this
requires the two partons to be produced at the same point in the
transverse space ($b=c$). As a result, this part is smaller than
$I^{(1)}_{AB}$, corresponding to independent production part , by
$\sim 1/A^{2/3}$ . It is important to recall that the independent
production part has been found without expansion in powers of
$1/A$, so that $I^{(2)}_{AB}$ fully represents the difference from
independent production.

%\section{Calculation of $I^{(2)}_{AB}$}
%\subsection{General structure}
To make the integrals over $r_1$ and $r_2$ convergent we change
unity in the last bracket to $\exp
(-T_A(b)F_A(w,0)-T_B(b)F_B(v,0))$ since this does not produce
terms which could contribute at $p,q\ne 0$ and present the
resulting cross-section in the form
\[
 I^{(2)}_{AB}(\beta,y_1,p,y_2,q)=P_A(v)P_B(w)\int d^2bT_A(b)T_B(b)
\int d^2r_1d^2r_2e^{ipr_1+iqr_2}\]
\[
\bigg(\int d^2l e^{il(r_1-r_2)}\frac{d\sigma(l,v,w)}{d^2l}
-\sigma^{tot}(v,w)\bigg)\]
\[
\bigg\{\bigg(e^{T_A(b)(F_A(w,r_2)-F_A(w,0))}-e^{-T_A(b)F_A(w,0)}\bigg)
\bigg(e^{T_b(b)(F_b(v,r_1)-F_B(v,0))}-e^{-T_B(b)F_B(v,0)}\bigg)
\]
\[
+e^{-T_A(b)F_A(w,0)}
\bigg(e^{T_b(b)(F_b(v,r_1)-F_B(v,0))}-e^{-T_B(b)F_B(v,0)}\bigg)
+e^{-T_B(b)F_B(v,0)}\]\beq
\bigg(e^{T_A(b)(F_A(w,r_2)-F_A(w,0))}-e^{-T_A(b)F_A(w,0)}\bigg)
+e^{-T_A(b)F_A(w,0)-T_B(b)F_B(v,0)}\bigg\}=
\sum_{i=1}^4I^{(2i)}_{AB} \eeq The most non-trivial correlations
are given by the first term. The simplest correlation is expressed
by the last term. Indeed, dropping terms proportional to
$\delta^2(p)$ or $\delta^2(q)$,
 \beq
I^{(24)}_{AB}=(2\pi)^4P_A(v)P_B(w)\delta^2(p+q)
\frac{d\sigma(p,v,w)}{d^2p} \int
d^2bT_A(b)T_B(b)e^{-T_A(b)F_A(w,0)-T_B(b)F_B(v,0)} \eeq and shows
back-to-back correlations. The second and third terms are
expressed via partonic and nuclear single inclusive
cross-sections:
 \beq
I^{(22)}_{AB}=(2\pi)^2P_B(w)\frac{d\sigma(q,v,w)}{d^2q} \int
d^2bT_B(b) e^{-T_A(b)F_A(w,0)}I_A(p+q,v,b)
 \eeq
  where we
defined the single cross-section at fixed $b$ as the integrand in
(20) (for the central collision, $\beta=0$) \beq
I_A(p,v,b)=P_A(v)T_A(b)\int d^2re^{ipr}
\bigg(e^{T_B(b)(F_b(v,r)-F_B(v,0))}-e^{-T_B(b)F_B(v,0)}\bigg) \eeq

The third term corresponds to $v,p\leftrightarrow w,q$ \beq
I^{(23)}_{AB}=(2\pi)^2P_A(v)\frac{d\sigma(p,v,w)}{d^2p} \int
d^2bT_A(b) e^{-T_B(b)F_B(v,0)}I_B(p+q,w,b) \eeq Finally in terms
of $I_{A,(B)}(p+q,v(w),b)$
\[
I^{(21)}_{AB}=\int d^2b \bigg(\int
d^2l\frac{d\sigma(l,v,w)}{d^2l}I_{A}(p+l,v,b) I_{B}(q-l,w,b)\]\beq
-\sigma^{tot}(v,w)I_A(p,v,b)I_B(q,w,b)\bigg) \eeq

\section{Correlations}

\label{corr}

We restrict ourselves to the study of the forward-backward
multiplicity correlations. Since the correlations are  in any case
small we may restrict ourselves to a linear dependence of the
average multiplicity in the backward window at a fixed
multiplicity in the forward window as a function of the latter;
\beq \label{lindep} \frac{\langle n_B\rangle_{n_F}}{\langle
n_B\rangle}(n_F) =a+b\,n_F \eeq Here $\langle n_B\rangle$ is the
overall average of the multiplicity in the backward window (at all
$n_F$). Thus defined coefficient $b$ shows the relative deviation
of the conditional average $\langle n_B\rangle_{n_F}$ when the
number of jets observed in the forward rapidity window $n_F$
changes by unity. It can be expressed via averages of linear and
bilinear products of multiplicities: \beq b=\frac{1}{\langle
n_B\rangle}\frac{\langle n_Bn_F\rangle -\langle n_B\rangle \langle
n_F\rangle }{\langle n^2_F\rangle -\langle n_F\rangle ^2} \label{bexpr} \eeq
%RK %%%%%
One gets this expression  by multiplying (\ref{lindep}) first by
$p(n_F)$ then by $n_F p(n_F)$, summing over $n_F$ and solving the
arising system of linear equations for $a$ and $b$.
%%%%%%%

%% RK %%%%%
To compute the mentioned bilinear products we first point out that
$n!$ in the denominator in the phase space volume for $n$
identical particles should refer only to particles within the same
phase space volume. For our problem it implies that identical
particles produced in different rapidity windows can be considered
as different (see Appendix for details).
%%%%%%%%%%%
%
%As a preliminary, we point out that particles emitted in different parts
%of the single particle  phase space may be considered as different,
%so that the factorials in the denominators should only refer to particles
%within the same phase space volume (see Appendix).
%For our problem it implies that we may consider particle produced
%in different rapidity windows as different.
This allows to immediately obtain simple expressions for the
squares of multiplicities coming from independent pair production
described by the product of single inclusive cross-sections.
Consider first emission of two particles into the forward rapidity
window. At a fixed  overall impact parameter $\beta$ we find \beq
\langle n_F\rangle=c\int\frac{dyd^2p}{(2\pi)^2}I_{A}(y,p) \eeq and
\beq \label{sigAAint} \langle
n_F(n_F-1)\rangle=c\int\frac{dy_1d^2p_1dy_2d^2p_2}{(2\pi)^4}
I_{AA}(\beta,y_1,p_1,y_2,p_2) \eeq where integrations over $y$ are
restricted to the forward rapidity window and
% RK
$c=1/(d^2\sigma_{AB}(\beta)/d^2\beta)$ where
$d^2\sigma_{AB}(\beta)/d^2\beta$ is the AB inelastic cross-section
at fixed impact parameter $\beta$. The latter is practically unity
for $\beta < R_A+R_B$.
%
% $c=1/\sigma_{AB}(\beta)\simeq 1$.
%
As shown in section~\ref{samedir} (Eq.~(\ref{sigAA})), in the
first approximation, the double inclusive cross-section in (45) is
just a product of two single inclusive ones. This gives a relation
\beq \langle n_F(n_F-1)\rangle=\frac{1}{c}\langle n_F\rangle^2
\eeq which, with $c\simeq 1$ leads to \beq \langle
n_F^2\rangle-\langle n_F\rangle^2=\langle n_F\rangle \label{samesidedisp}\eeq
(effectively the distribution seems to be Poissonian). Note that
the dispersion is different from zero. This is the reason why for
the production of the pair into the same (forward) rapidity window
we can limit to the first approximation in powers of $1/A$ or
$1/B$.

Passing to the emission of jets into different rapidity windows,
we may consider the jets different. So instead of (\ref{sigAAint})
we shall find \beq \langle
n_Fn_B\rangle=c\int\frac{dy_1d^2pdy_2d^2q}{(2\pi)^4}
I_{AB}(\beta,y_1,p,y_2,q) \eeq From independent production (part
$I_{AB}^{(1)}$) we shall get just the product of average
multiplicities, so that \beq \langle n_Fn_B\rangle-\langle
n_F\rangle\langle n_B\rangle=
\int\frac{dy_1d^2pdy_2d^2q}{(2\pi)^4}
I_{AB}^{(2)}(\beta,y_1,p,y_2,q) \eeq where we used $c\simeq 1$.

This gives for the correlation coefficient $b$ \beq
\label{corcoef} b=\frac{\int dy_1d^2pdy_2d^2q
I_{AB}^{(2)}(\beta,y_1,p,y_2,q)} {\int dy_1d^2p I_{A}(y_1,p) \int
dy_2d^2q I_{A}(y_2,q)} \eeq where integrals over $y_1,p$ and
$y_2,q$ go over the forward and backward rapidity windows,
respectively. We see that in the numerator of the expression for
the correlation coefficient the leading terms in powers   of $1/A$
and $1/B$ cancel and only  subleading terms of the relative order
$1/A^{2/3}$ or $1/B^{2/3}$ remain. This means that in any case
forward-backward multiplicity correlations at a fixed $\beta$ have
the order  $1/A^{2/3}$or $1/B^{2/3}$. So for their observation
collision of comparatively light nuclei is preferable. To
calculate the correlations one has to evaluate the integrals in
(\ref{corcoef}).

Our definition of the correlation coefficient differs from the
conventional one which reads $b_{\rm 0}={\rm
Cov}(n_B,n_F)/\sqrt{D(n_B)D(n_F)}$ and for equal dispersions of
forward and backward multiplicities is $b_{\rm 0}=(\langle n_F
n_B\rangle-\langle n_F\rangle\langle n_B\rangle)/ (\langle n_F^2
\rangle - \langle n_F\rangle^2)$. As one can see comparing $b_{\rm
0}$ to (\ref{bexpr}) the difference is by factor $\langle
n_B^2\rangle - \langle n_B \rangle\langle n_B \rangle$. As this
factor is proportional to the backward rapidity window width
$\Delta y_2$ (see (\ref{samesidedisp})), the correlation
coefficient (\ref{bexpr}) has an advantage compared to $b_{\rm 0}$
having no explicit dependence on the chosen rapidity intervals.
However the value of correlation coefficient $b$ is not limited
from above by unity as for $b_{\rm 0}$.

\section{Numerical results}

\label{numres}

In the general case calculation of (\ref{corcoef}) presents a
formidable numerical task. To simplify it we limit ourselves to
central collisions of identical nuclei ($A=B$ and $\beta=0$). We
also assume the two rapidity windows narrow in rapidity and
magnitudes of the transverse momenta $|p|$ and $|q|$ centered
around $y_1,p$ and $y_2,q$, so that in (\ref{corcoef}) the
inclusive cross-sections can be taken out of the integrals in
$y_{1,2}$ and $|p|$ and $|q|$ at these points.
% change!
As to the azimuthal angle $\phi$ between the jet momenta $\bf p$ and $\bf q$
it may be chosen differently in the experimental setup. If one takes
into account pairs of jets with an arbitrary $\phi$ then
after
integration over the angles we will get a correlation coefficient
depending only on the chosen $y_1,p$ and $y_2,q$:
 \beq
b(y_1,p,y_2,q)=\frac{1}{\pi} \frac{\int d\phi
I_{AB}^{(2)}(\beta=0,y_1,p,y_2,q,\phi)} {I_{A}(y_1,p)I_{B}(y_2,q)}
\label{corcoefcomp}
\eeq
We restricted our calculations to the case
$y_1=y_2$ and $p=q$.

Note that the elementary processes taken into account in
$I^{(2)}_{AB}$ formally include the contribution from a single hard
rescattering  accompanied by multiple soft rescatterings.  Physically it
corresponds to smearing of the lowest order back-to-back correlation
by soft interactions both before annd afterwards. In our formalism
soft interactions are suppressed by a cutoff in the elementary parton-parton
cross-section. However one can see that at large jet transverse momenta
the total correlations become dominated by the above mentioned contribution
with soft rescatterings taken at the transverred momenta of the order of
the cutoff. Obviously such contribution cannot be desribed correctly in our
perturbative formalism. To eleminate the effect of
soft-smeared back-to-back correlations, especially pronounced at high
momenta, we choose to restrict the experimental range of azimuthal angles
$\phi$ between the jets, excluding from it angles close to the back-to-back
configuration. We introduce
%%%%%%  Veto angle origin and imposed modificaltion of the coefficient  %%%%%%%
%Now the angular structure of the correlated part $I^{(2)}_{AB}$ is
%worth discussing. Among all $I^{(2)}_{AB}$ contains contributions
%from multiple soft rescatterings of the observed partons off the
%unobserved ones, and one single hard rescattering of the observed
%off each other. This corresponds to almost all the transverse
%momentum of the observed partons being delivered by their single
%mutual scattering. This remnant of the back to back correlations
%dominates at the asimuthal angle between jets close to $\pi$ and
%is especially large at high transverse momenta values. To
%eliminate this trivial contribution
a ``veto angle'' for the final momenta, that is we demand that the azimuthal
angle $\phi$ should not be  larger than $\pi-\phi_{veto}/2$. This
excludes the undesirable interval of $\phi$ around $\pi$ of
length $\phi_{veto}$. The value
$\pi-\phi_{veto}/2$ then serves as the upper limit of integration
in the numerator of (\ref{corcoefcomp}) and the denominator is
modified by a factor of $(\pi-\phi_{veto}/2)/\pi$.

For the partonic  cross-sections we have taken the same expression
as in~\cite{bra}. Namely we assumed the effective partonic
distributions with the gluon-gluon cross-section as a dynamical
input. The lowest order cross-section was multiplied by the
so-called $K$ factor to take into account higher order
corrections. The infrared regularization was realized by a
non-zero gluon mass $p_0$. The scale in the partonic distributions
was taken as $p_0^2+p_1^2$ where $p_1$ was the transferred
transverse momentum.

We studied S-S and Pb-Pb colisions at two c.m. energies 200 and
6000 GeV. In accordance with \cite{trel2} we set $p_0=1$~GeV and
$K=1.04$ at 200~GeV and $p_0=2$~GeV and $K=2$ at 6000~GeV. In our
calculations we take $\phi_{veto}=\pi/6$. We also present results
for 200~GeV$\cdot A$ Cu-Cu and Au-Au collisions and for
200~GeV$\cdot A$ S-S with $\phi_{veto}=\pi/3$.

In figures 1 and 2 we show the correlation coefficient $b$ as a
function of $p$ at $y_1=y_2=1,2$ and 3 for S-S and Pb-Pb
collisions at c.m. energy 200~GeV. Figures 3 and 4 show the same
correlation coefficients at 200~GeV for Cu and Au nuclei and
figures 5 and 6 represent the correlation coefficient for S and Pb
at 6000~GeV.

\begin{figure}[!htbp]
\centering
\includegraphics[angle=-90,width=12cm]{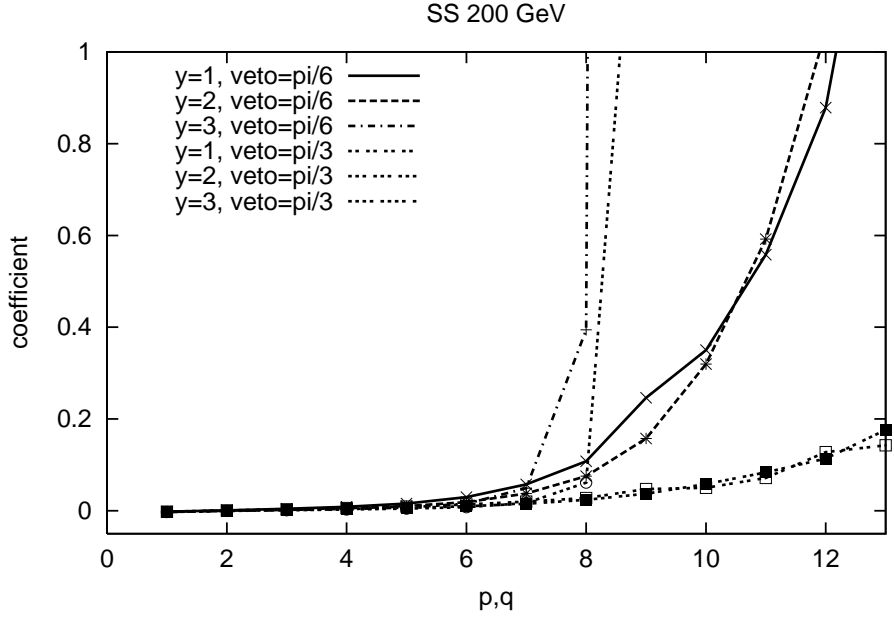}
\caption{Correlation coefficient for sulphur-sulphur collisions at
$\sqrt s = 200$~GeV.}
\end{figure}

\begin{figure}[!htbp]
\centering
\includegraphics[angle=-90,width=12cm]{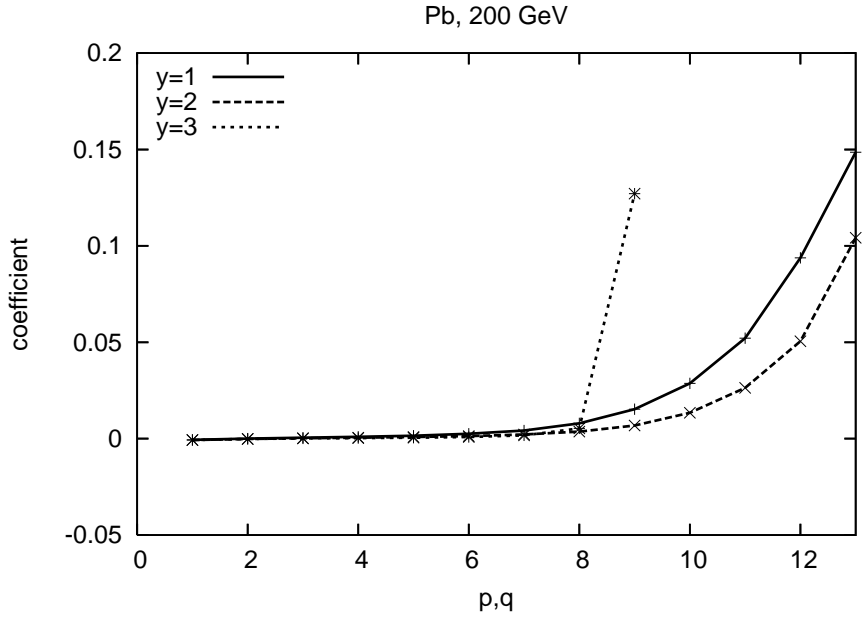}
\caption{Correlation coefficient for lead-lead collisions at
$\sqrt s = 200$~GeV.}

\end{figure}

\begin{figure}[!htbp]
\centering
\includegraphics[angle=-90,width=12cm]{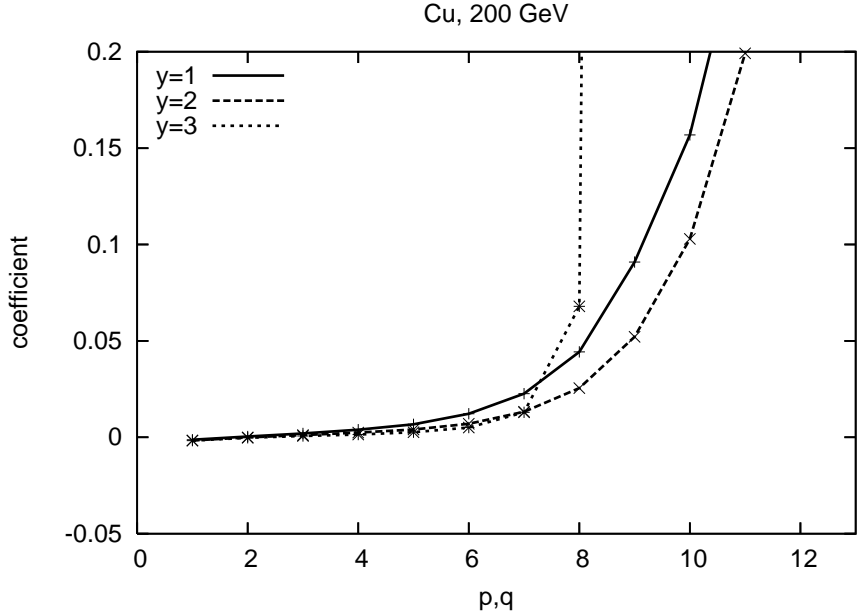}
\caption{Correlation coefficient for copper-copper collisions at
$\sqrt s = 200$~GeV.}
\end{figure}

\begin{figure}[!htbp]
\centering
\includegraphics[angle=-90,width=12cm]{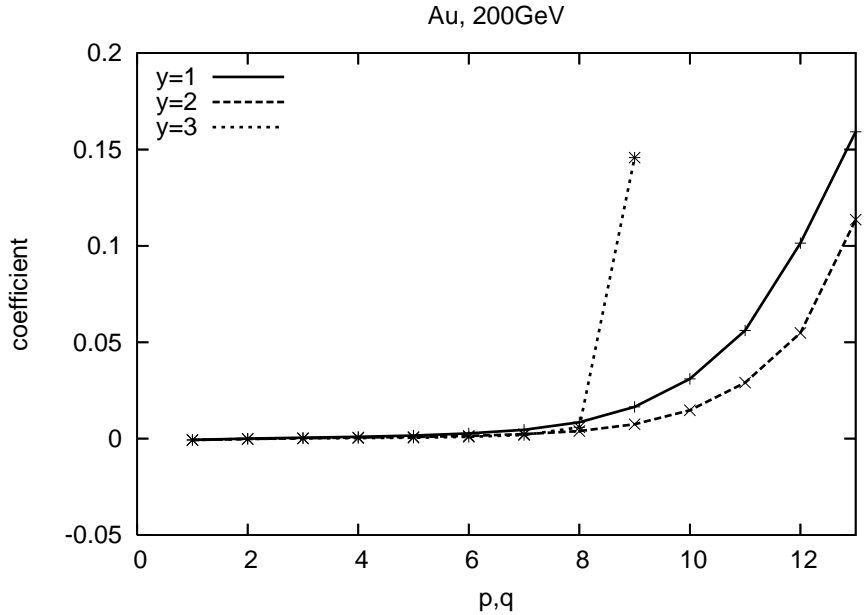}
\caption{Correlation coefficient for gold-gold collisions at
$\sqrt s = 200$~GeV.}

\end{figure}

\begin{figure}[!htbp]
\centering
\includegraphics[angle=-90,width=12cm]{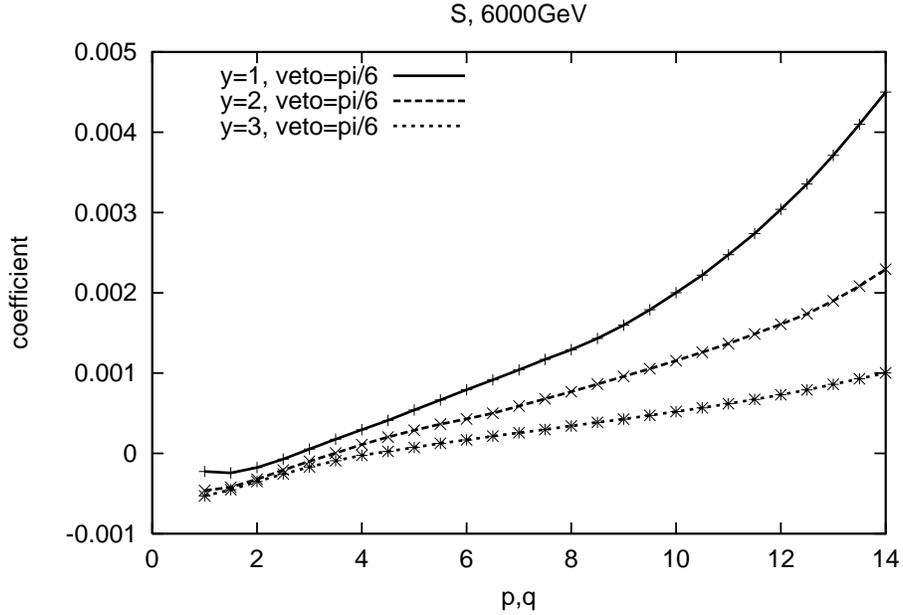}
\caption{Correlation coefficient for sulphur, $\sqrt s =
6000$~GeV.}
\end{figure}

\begin{figure}[!htbp]
\centering
\includegraphics[angle=-90,width=12cm]{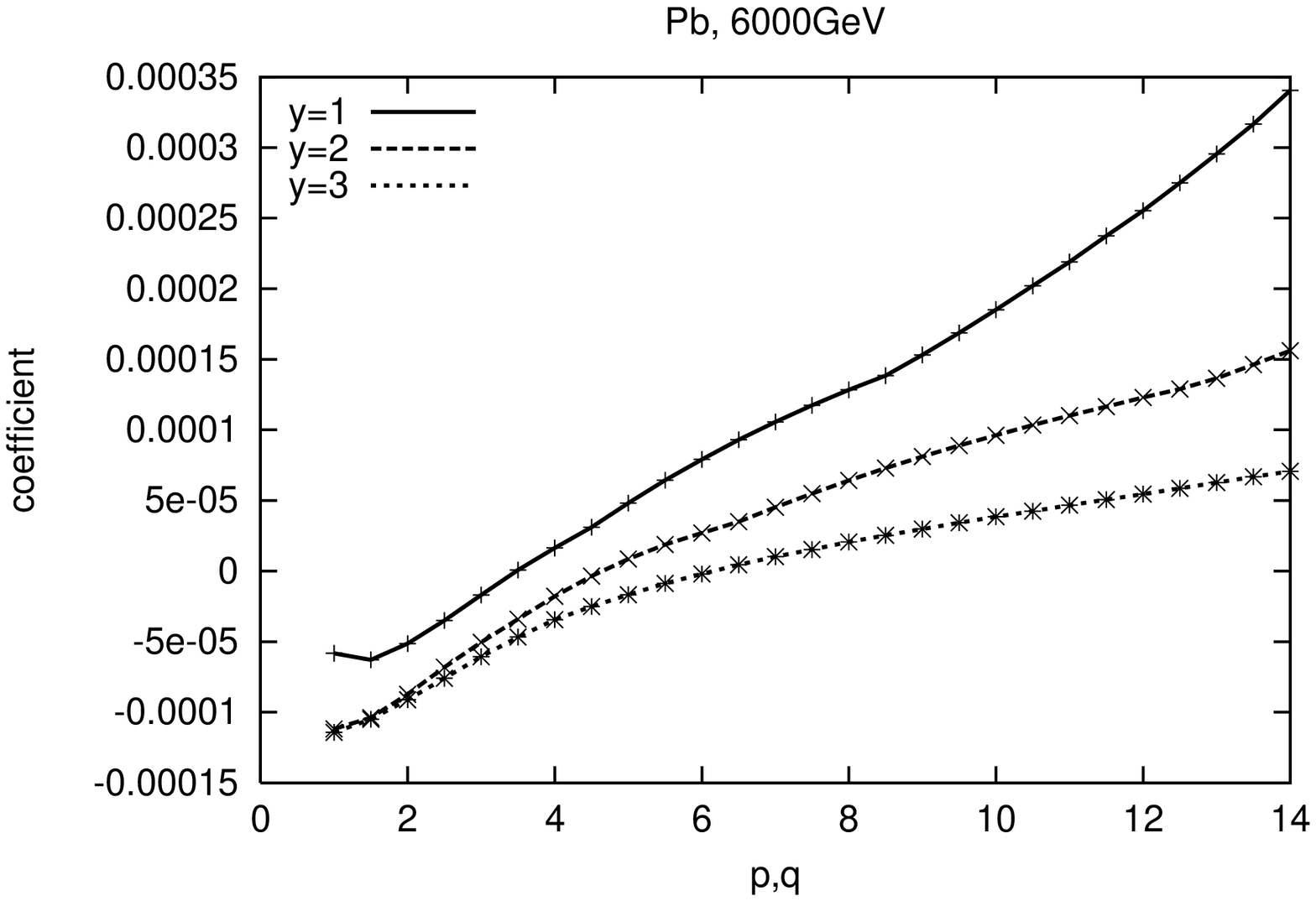}
\caption{Correlation coefficient for lead, $\sqrt s = 6000$~GeV.}
\end{figure}

%%%%%%%%%%%%%%%%%%%%%%%%%%%%%%%%%%%%%%%%%%%%%%%%%%%%%%%%%%%%%%%%%%%%%
 The found correlation coefficients drop with atomic numbers and
energy and grow with jet momenta. For Su-Su collisions at 200 GeV
they have a sharp rise for momenta around 7$\div$8 GeV/c and
rapidity $y_B=y_F=3$. One of the reasons for this was already
discussed in this section, another is due to to kinematics: the
single inclusive cross-sections in the denominator of (43) tend to
zero at the limits of the phase space volume. As expected, for
Pb-Pb collisions the correlation coefficients are an order of
magnitude smaller than for S-S collisions.

\section{Conclusions}

\label{concl}

Our calculations show that in spite of the fact that in heavy
nuclei  long range multiplicity correlations between jets are
small, of the order $A^{-2/3}$ for identical nuclei, they are of
observable magnitude if nuclei are not too heavy. They also
visibly grow with the transverse momentum and for S-S collisions
at 200~GeV. %they reach the level of 15$\div$20  at momenta of both
%jets around 7 GeV/$c$.
For transverse momenta of jets of about $10\div12$~GeV/$c$ they
stay large ($10\div20$\%) even if the back-to back correlations
are totally excluded by $\pi/3$ veto angle. Their observation and
measurement for non-zero veto angles will
%%RK
favor a hypothesis
%be a clear indication
that multiple hard collisions indeed occur before the
fragmentation of jets into hadrons and are described by the
perturbative QCD mechanism. However at the supposed LHC energies
of 6~TeV these correlations are strongly suppressed reaching 0.5\%
for S-S collisions at jet momenta of about 15~GeV/$c$.

Of course we understand that other effects may somewhat change our
predictions based on the simple Glauber approach. Among them the
most prominent is quenching of jets as they propagate through the
nuclear medium. Also if the jet energy happens to  be
insufficiently high  the formation length may shorten to allow for
the jets hadronization inside the nucleus, which will spoil our
simple picture of hard rescattering.
%%%% RK
Another important point is the jet fragmentation which in general
can lead to correlation pattern for hadrons different form that
for partons.
%%%%
These questions are under our consideration at present. In any
case these other effects will produce certain corrections to the
results presented in this paper, which thus may serve as a natural
starting point.

\section{Acknowledgements}
The authors are grateful to Dr. V.M.\,Suslov for his advice in the
computational techinques. M.A.B. is  thankful to the North
Carolina Central University for hospitality and support. This work
was  supported by grants of Education Ministry of Russia RNP
2.1.1.1112 and RFFI of Russia 06-02-16115a. The work of R.S.K. was
supported by INTAS Nr.~05-112-5031.

\section{Appendix.  Identical particles in different parts of
the phase volume (related to Eq. (47)}
    Consider two particles characterized, say,
by their rapidities only  $y_1$ and $y_2$ which take values in the
interval [0,1].
% RK
The integrals over intermediate state of two identical particles,
between which no distinction is made, in general  have the form:
%%%
%The integrals over intermediate particle states then have the form
  \beq
  I=\int _0^1\frac{dy_1dy_2}{2!}f(y_1,y_2)
  \eeq
where $f(y_1,y_2)$ is some function symmetric in $y_1$ and $y_2$.
Now we split the single particle phase volume  in two, say
[0,1]=[0,1/2]+[1/2,1] and assume that particle with its rapidity
in [0,1/2] is particle 1 and that with its rapidity in [1/2,1] is
particle 2. Now instead of a single intermediate state we have 3
different ones: with two particles 1 (state A), with two particles
2 (state B) and with one particle 1 and one 2 (state C). The same
integral will now be given by a sum
  \beq
  I_A +I_B+I_C=\int_0^{1/2}\frac{dy_1dy_2}{2!}f(y_1,y_2)+
  \int_{1/2}^{1}\frac{dy_1dy_2}{2!}f(y_1,y_2)
  +\int_0^{1/2}dy_1\int_{1/2}^1dy_2f(y_1,y_2)
  \eeq
As we see this sum is equal to $I$ exactly. So considering
particles in [0,1/2] and [1/2,1] as different gives the correct
result.

\end{document}